\documentclass[sigconf]{acmart}

\usepackage{verbatim}
\usepackage{multirow}

\AtBeginDocument{%
  }

\copyrightyear{2024} 
\acmYear{2024} 
\setcopyright{acmlicensed}
\acmConference[ASE '24]{39th IEEE/ACM International Conference on Automated Software Engineering}{October 27-November 1, 2024}{Sacramento, CA, USA}
\acmBooktitle{39th IEEE/ACM International Conference on Automated Software Engineering (ASE '24), October 27-November 1, 2024, Sacramento, CA, USA}
\acmDOI{10.1145/3691620.3695537}
\acmISBN{979-8-4007-1248-7/24/10}

\usepackage{enumitem}
\begin{document}

\title{Enhancing Automated Program Repair with Solution Design}



\author{Jiuang Zhao}
\authornote{These authors contributed equally to this work.}
\email{zja1999@buaa.edu.cn}
\affiliation{%
  \institution{School of Computer Science and Engineering, Beihang University}
  \city{Beijing}
  \country{China}
}

\author{Donghao Yang}
\authornotemark[1]
\email{yangdonghao@buaa.edu.cn}
\affiliation{%
  \institution{School of Computer Science and Engineering, Beihang University}
  \city{Beijing}
  \country{China}
}

\author{Li Zhang}
\email{lily@buaa.edu.cn}
\affiliation{%
  \institution{CCSE, Beihang University}
  \city{Beijing}
  \country{China}
  }

\author{Xiaoli Lian}
\authornote{Corresponding author.}
\email{lianxiaoli@buaa.edu.cn}
\affiliation{%
  \institution{CCSE, Beihang University}
  \city{Beijing}
  \country{China}
}

\author{Zitian Yang}
\email{yangzitian@buaa.edu.cn}
\affiliation{%
  \institution{School of Software, Beihang University}
  \city{Beijing}
  \country{China}
}

\author{Fang Liu}
\email{fangliu@buaa.edu.cn}
\affiliation{%
  \institution{CCSE, Beihang University}
  \city{Beijing}
  \country{China}
}


\begin{abstract}

Automatic Program Repair (APR) endeavors to autonomously rectify issues within specific projects, which generally encompasses three categories of tasks: bug resolution, new feature development, and feature enhancement. Despite extensive research proposing various methodologies, their efficacy in addressing real issues remains unsatisfactory. It's worth noting that, typically, engineers have design rationales (DR) on solution— planed solutions and a set of underlying reasons—\emph{before} they start patching code. In open-source projects, these DRs are frequently captured in issue logs through project management tools like Jira. This raises a compelling question: \emph{How can we leverage DR scattered across the issue logs to efficiently enhance APR?}

To investigate this premise, we introduce \textit{DRCodePilot}, an approach designed to augment GPT-4-Turbo's APR capabilities by incorporating DR into the prompt instruction. Furthermore, given GPT-4's constraints in fully grasping the broader project context and occasional shortcomings in generating precise identifiers, we have devised a feedback-based self-reflective framework, in which we prompt GPT-4 to reconsider and refine its outputs by referencing a provided patch and suggested identifiers.
We have established a benchmark comprising 938 issue-patch pairs sourced from two open-source repositories hosted on GitHub and Jira. Our experimental results are impressive: \textit{DRCodePilot} achieves a full-match ratio that is a remarkable 4.7x higher than when GPT-4 is utilized directly. Additionally, the CodeBLEU scores also exhibit promising enhancements. Moreover, our findings reveal that the standalone application of DR can yield promising increase in the full-match ratio across CodeLlama, GPT-3.5, and GPT-4 within our benchmark suite.
We believe that our \textit{DRCodePilot} initiative heralds a novel human-in-the-loop avenue for advancing the field of APR.

\end{abstract}





\maketitle

\section{Introduction}
\label{introduction}
    
    Due to persistent tasks such as fixing bugs, introducing new features, and improving existing ones, software maintenance consistently ranks as the most time-consuming phase in the software life cycle, even accounting for up to 90\% of total costs \cite{rashid2009gauging}. This phase is particularly challenging because it requires maintainers to possess a deep understanding of the large and complex codebase, to analyze, design, and implement modifications without introducing new issues. To expedite this process, various Automated Program Repair (APR) techniques have been proposed to lessen the burden \cite{xia2023_automated_program_repair_in_the_era_of_large_pre-trained,A_Survey_on_Automated_Program_Repair_Techniques_Kai2023}. However, their effectiveness in addressing real-world problems remains suboptimal \cite{jimenez2024swebench,Noda2020HowEffective}.
    
\begin{figure}[!htbp]
        \centering   \includegraphics[width=0.95\linewidth]{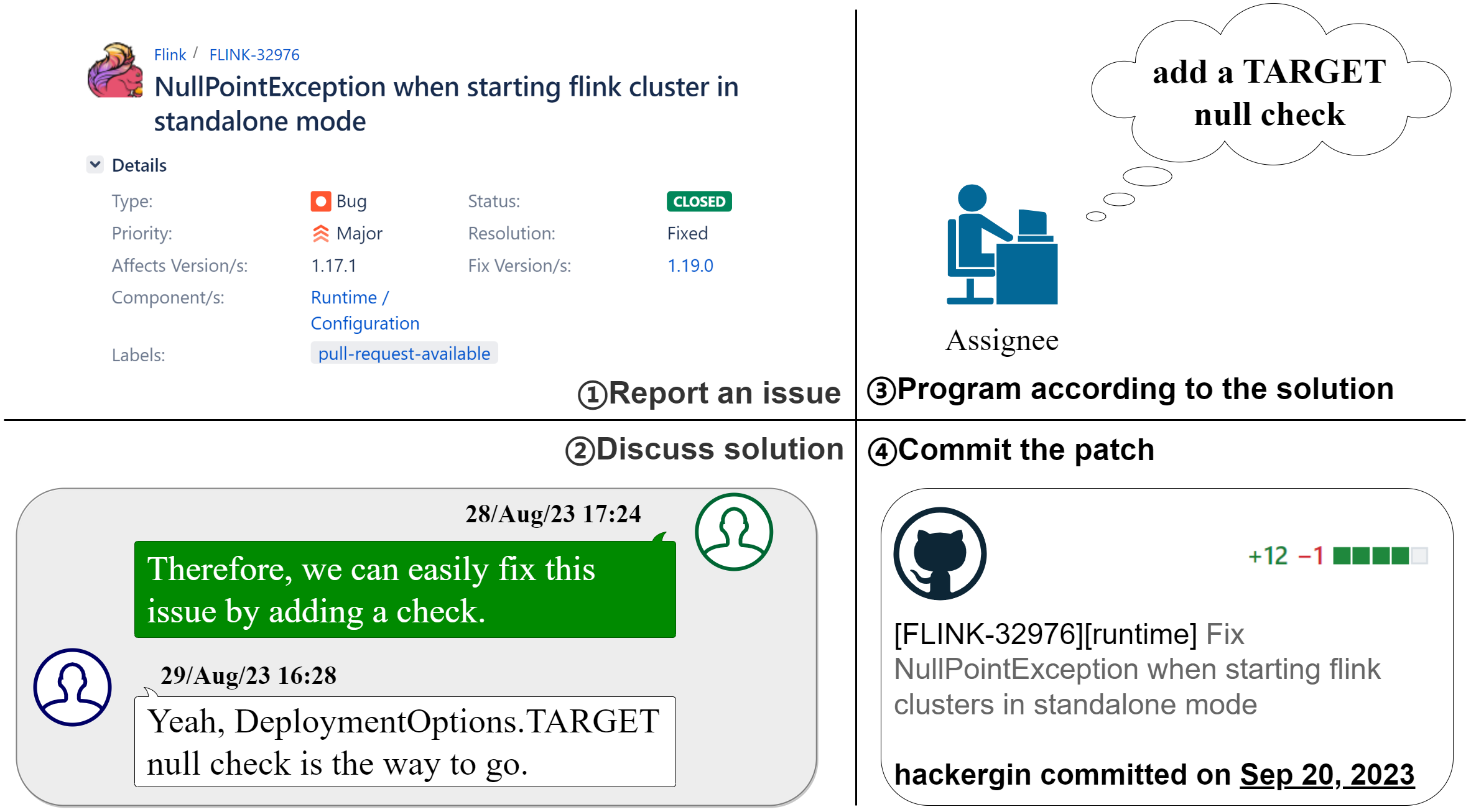}
        \caption{Four Stages of Issue Resolving in Open-source Community(Jira + GitHub): With One Example of FLINK-32976.}
        \Description{}
        \label{fig:Flink-32976}
    \end{figure}

    Let us reflect on the human process of software maintenance. In practice, engineers typically have an idea of the solution along with its associated considerations before beginning the actual code patching process. Within the open-source community, project management tools such as Jira are commonly employed to document this patching workflow. We depict this with the case study of addressing issue FLINK-32976, as presented in Fig. \ref{fig:Flink-32976}. The process is initiated by an issue report submitted by either engineers or users. Subsequently, contributors engage in discussions about the issue, possible solutions, and their respective advantages and drawbacks. An assignee then codes the solution, informed by these discussions. The final step involves the engineers submitting the patch to GitHub via a pull request.

    It is evident that solutions and their accompanying arguments in the discussion, collectively termed the design rationale (DR) \cite{burge2008software,Alkadhi2017_RationaleinDevelopmentChatMessages,Using_Rationale_to_Support_wang_978-1-60558-967-1,Learning_the_"Whys"_Discovering_Design_Rationale_Liang,ARationale-basedArchitectureModel_Tang_0164-1212,A_New_Design_Rationale_Representation_Model_for_Rationale_Mining_Ying_1530_9827,Design_Knowledge_and_Design_Rationale:A_Framework_for_Representation_Capture_and_Use_ThomasR},  play an indispensable role in manual program maintenance \cite{Burge2003Rationale_SupportforMaintenanceofLargeScaleSystems}. They furnish engineers with guiding principles for developing patch, thus reducing the incidence of flawed fixes \cite{Zilouchian2012Consensus}. This observation prompts us to inquire: \emph{How to effectively incorporate these design rationales to enhance the performance of current APR?}

    In our examination of the current state of APR research, we observed limited utilization of design rationale from discussions. Semantic search methods primarily focus on code similarity search, leveraging existing code repositories to find fixes that previously worked in similar contexts\cite{GenProg_a_Generic_Method_for_Automatic_Software_Repair_Le2012,Shaping_Program_Repair_Space_with_Existing_Patches_and_Similar_Code_Jiang2018}. Semantic-constraint \cite{nopol2016xuan,On_the_efficiency_of_test_suite_based_program_repair_liu2020} and pattern-based \cite{kim2013_Automatic_patch-generation_learned_from_human-written_patches,le2016_historydrivenprogramrepair,koyuncu2019ifixr_Bugreportdrivenprogramrepair,liu2019tbar} methods further optimize the search process through manually defined rules or templates. These methods are ineffective in leveraging higher-level repair knowledge from human experience \cite{A_Survey_on_Automated_Program_Repair_Techniques_Kai2023}. Learning-based models \cite{marquez2018empirical_studyofscalabilityframeworksinopensourcemicroservices-basedsystems,white2019sorting_transforming_programrepairingredients,li2020dlfix_Context-basedcodetransformation,lutellier2020coconut_combiningcontext-awareneuraltranslation,A_Survey_on_Automated_Program_Repair_Techniques_Kai2023} learn empirical knowledge from a large number of defect repair samples to guide the repair process. Existing work \cite{DBLP:conf/emnlp/PanthaplackelGL22} has trained deep learning models to generate code patches by adhering to high-level guidelines (e.g. solution description summarized from discussions); however, the models' performance has been limited. This may be attributed to three key factors: firstly, the inherent noise and complexity present in interleaved conversational data \cite{AnalysisandDetectionofInformationTypesofOpenSourceSoftwareIssueDiscussions}, making it challenging to obtain high-quality design rationale from discussions \cite{panthaplackel-etal-2022-learning}; secondly, the intricate process of collaborative problem-solving represented by design rationale \cite{Zilouchian2012Consensus}, which demands strong reasoning abilities \cite{CHANDRASEGARAN2013204knowledgeRepresentation}; and lastly, the models may lack the requisite background knowledge to properly articulate the technical solutions under discussion \cite{Allamanis2014coding_conventions,tian2024codehalu}.
    
    We note that work have emerged for efficiently mining design rationale from issue logs \cite{zhao2024novel}, providing a solid foundation for leveraging design rationale. Moreover, large language models (such as GPT-4-turbo \cite{bubeck2023sparksgpt-4}) have the potential to better bridge the gap between abstract design rationale and concrete source code patches \cite{xia2023_automated_program_repair_in_the_era_of_large_pre-trained} : Some work has demonstrated that LLMs can integrate different types of software artifacts (e.g., bug reports) to better repair programs \cite{zhang2024autocoderover,motwani2023better}. Therefore, we aim to explore methods for efficiently integrating design rationale into APR based on LLM and existing tools. The previously analyzed challenges in utilizing design rationale may lead to LLMs experiencing reasoning failures and code hallucinations \cite{tian2024codehalu}. Fortunately, recent research has shown that LLMs can better handle complex problems through various types of feedback and self-correction mechanisms \cite{ji2023towards_Mitigating_Hallucination,shinn2023reflexion}. In light of this, we introduce a feedback-based self-reflection framework to better empower LLM to apply design rationales.
    
    Specifically, we introduce \textit{DRCodePilot}, an approach that harnesses design rationale to drive automated patch generation. Initially, DRs are derived by extracting and pairing issue solutions with their corresponding arguments, leveraging the DRMiner tool\cite{zhao2024novel}. Then GPT-4 pinpoints defective segments and generates patches by considering all DRs. For refining the patches, \textit{DRCodePilot} collects feedback that encapsulates project-specific knowledge from two distinct sources:(1) reference patches generated by a fine-tuned CodeT5P \cite{wang2023codet5p}, and (2) identifier replacement suggestions through a retrieval technique. Armed with this feedback, GPT-4 reflects on its initial answer, reasoning out final, refined patches.

    To conduct an evaluation, we created a benchmark consisting of 938 issue-patch pairs by correlating Jira issues with their respective GitHub commits from two actively developed open-source projects: Flink and Solr. We selected five advanced code LLMs as baselines, including CodeLlama \cite{roziere2023codellama}, StarCoder2 \cite{lozhkov2024_starcoder2}, CodeShell \cite{xie2024codeshell}, GPT-3.5-Turbo, and GPT-4-Turbo (simplified as GPT-3.5 and GPT-4). 
    
    The results of our experiments demonstrate the superior performance of our \textit{DRCodePilot}. Notably, in terms of the number of full-match patches (those identical to the gold patches), our model achieved \emph{109 full matches out of 714 samples} in the Flink dataset and \emph{18 out of 224} in the Solr dataset. This significantly surpasses the best-performing baseline model, GPT-4, which only managed \emph{23 and 5 full matches} respectively, making our model 4.7 and 3.6 times more effective. Furthermore, \textit{DRCodePilot} improved CodeBLEU scores by 5.4\% and 3.9\% over GPT-4. Our findings also highlight the beneficial influence of DR, alongside reference patch and identifier feedback, on the quality of generated patches. Despite the imperfections of automated DR extraction by DRMiner, the sole application of these extracted DRs has proven to obviously enhance the full-match ratio across all baseline models impressively. We also showcase that patch quality can be further elevated with improvements in DR quality. 
    
    Our main contributions are outlined as follows:

    \begin{itemize}[leftmargin=2mm]
    \item \textbf{Framework}: We introduce \textit{DRCodePilot}, a novel feedback-based self-reflective framework that mimics human process in software maintenance. It begins by generating initial patches grounded in design rationale and refines them based on the feedback of reference patches and identifier recommendations. To our knowledge, \textit{DRCodePilot} is the pioneering effort to merge design rationale with APR techniques.
    
    \item \textbf{Data}: We collect up to 938 issue-patch pairs from two open-source projects and construct a dataset. We public the dataset and source code of \textit{DRCodePilot}\footnote{\url{https://figshare.com/s/82ed8e86e88d3268b4c1}} to facilitate the replication of our study and its application in extensive contexts.
        
    \item \textbf{Evaluation}: We conduct experiments on our dataset to show that DRCodePilot achieves substantial improvements across different projects, especially a much higher full-match ratio. Ablation experiments further confirms the significant potential of applying design rationales in LLM-based software maintenance.
    \end{itemize}

\section{Motivating example}
\label{sec:background}

\begin{figure}[!htbp]
        \centering
        \includegraphics[ trim = {0, 0, 0, 0}, clip, width=0.98\linewidth]{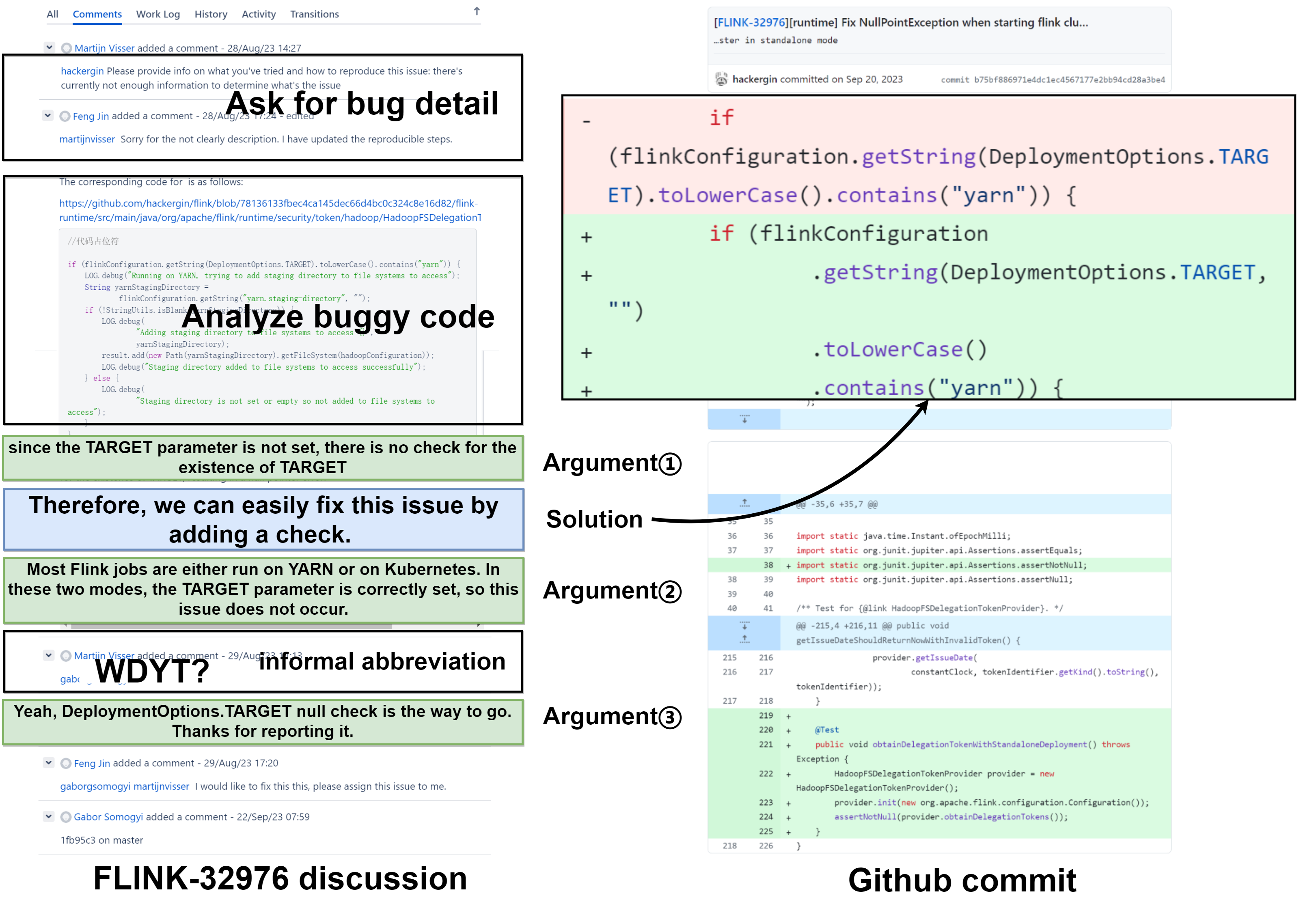}
        \caption{Classification of Issue Comments and Mapping from Solution to Patch for FLINK-32976.}
        \Description{}
        \label{fig:motivating}
    \end{figure}

    In this section, we illustrate the potential impact of design rationale on program maintenance using a simple issue example, FLINK-32976\footnote{\url{https://issues.apache.org/jira/browse/FLINK-32976}} in Jira, as depicted in Figure \ref{fig:motivating}.

    The discussion of FLINK-32976 centers on a null pointer exception that occurs when starting a Flink standalone cluster with Hadoop configuration. The title and description explain the issue. During the comments zone, one participant \emph{Hackergin} provided reproduction steps and the buggy code. He identified the root cause: the absence of a TARGET parameter check when the standalone cluster is initiated, leading to the null pointer exception. He also suggested one solution: adding a null check for the TARGET parameter, which all participants agreed was the best approach.
    
    After identifying the solution, the assignee referenced the design rationales discussed to shape the final code implementation. However, there is a notable gap between design rationale and actual code implementation, requiring developers to have an in-depth understanding of the codebase. For instance, while the design rationale indicated the need to check the TARGET parameter, in practice, developers ensured robustness by setting an empty string as the default value.

    We observed that despite the issue being classified as major and the solution has already been determined, it still took the maintainer 20 days to commit the changes. Such delays can introduce several risks. Automating adherence to these design rationales and generating patches could greatly improve efficiency—a capability not currently supported by existing APR techniques.

    Moreover, while the example depicted has only 5 comments, more complex scenarios exist, such as FLINK-34007\footnote{\url{https://issues.apache.org/jira/browse/FLINK-34007}}, which involved 82 detailed comments, debated three main solutions, the involving steps and arguments. In these complex issues, even advanced LLMs suffer from input length limits by injecting the whole issue logs, and the noisy nature of logs may hinder the accurate capture of design information \cite{liu2023lost_in_the_middle}. Furthermore, due to the lack of project-specific background knowledge, LLMs might produce erroneous outputs, leading to inefficiencies.

    In summary, design rationales are a source of valuable guidance for engineers as they resolve practical issues. However, these rationales are often abstract and can sometimes be conflicting when multiple are presented for a single issue, necessitating a deep understanding of the project's intricacies by the engineers. Additionally, design rationales offer only high-level recommendations, granting engineers the discretion to determine if and how they should be applied. In effect, the challenge lies in harnessing these rationales automatically and effectively. 

    This is where our research comes into play. We aim to leverage the code comprehension and reasoning capability of large language models, exploring the extent to which DR can aid in their ability to automatically repair issues.

\section{Approach}
\label{ourApproach}

In this section, we discuss the design of \textit{DRCodePilot}. \textit{DRCodePilot} is designed to work in a realistic software development lifecycle, in which users submit issue reports to a software repository for bug fixing, feature addition or improvement, and the project maintainers have discussions about the issue before craft a patch to resolve it. 

The \textit{DRCodePilot} methodology unfolds in five strategic phases: 1) \textit{Design Rationale Acquisition} (Section \ref{subsec:DRAcquisition}), where we replicate the existing work \cite{zhao2024novel} to mine solutions and corresponding reasons for specific issues from Jira logs; 2) \textit{Defective Segment Location and Draft Patch Generation} (Section \ref{subsec:draftGeneration}), pinpointing flawed code areas and crafting initial patches via GPT-4, leveraging the gathered design rationales; 3) \textit{Reference Draft Generation} (Section \ref{subsec:referenceGeneration}), employing a fine-tuned \emph{CodeT5P} model against the full project repository to generate benchmarks, addressing GPT-4's long-context limitations; 4) \textit{Identifier Recommendation} (Section \ref{subsec:identifierRecommendation}), offering alternatives for the potential unsuitable identifiers in auto-generated segments; 5) \textit{Final Patch Generation} (Section \ref{subsec:finalPathGeneration}), directing GPT-4 to improve initial drafts based on reference patches and identifier suggestions.

    \begin{figure*}[htb!]
        \centering
        \includegraphics[ trim = {0, 0, 0, 0}, clip, width=1\textwidth]{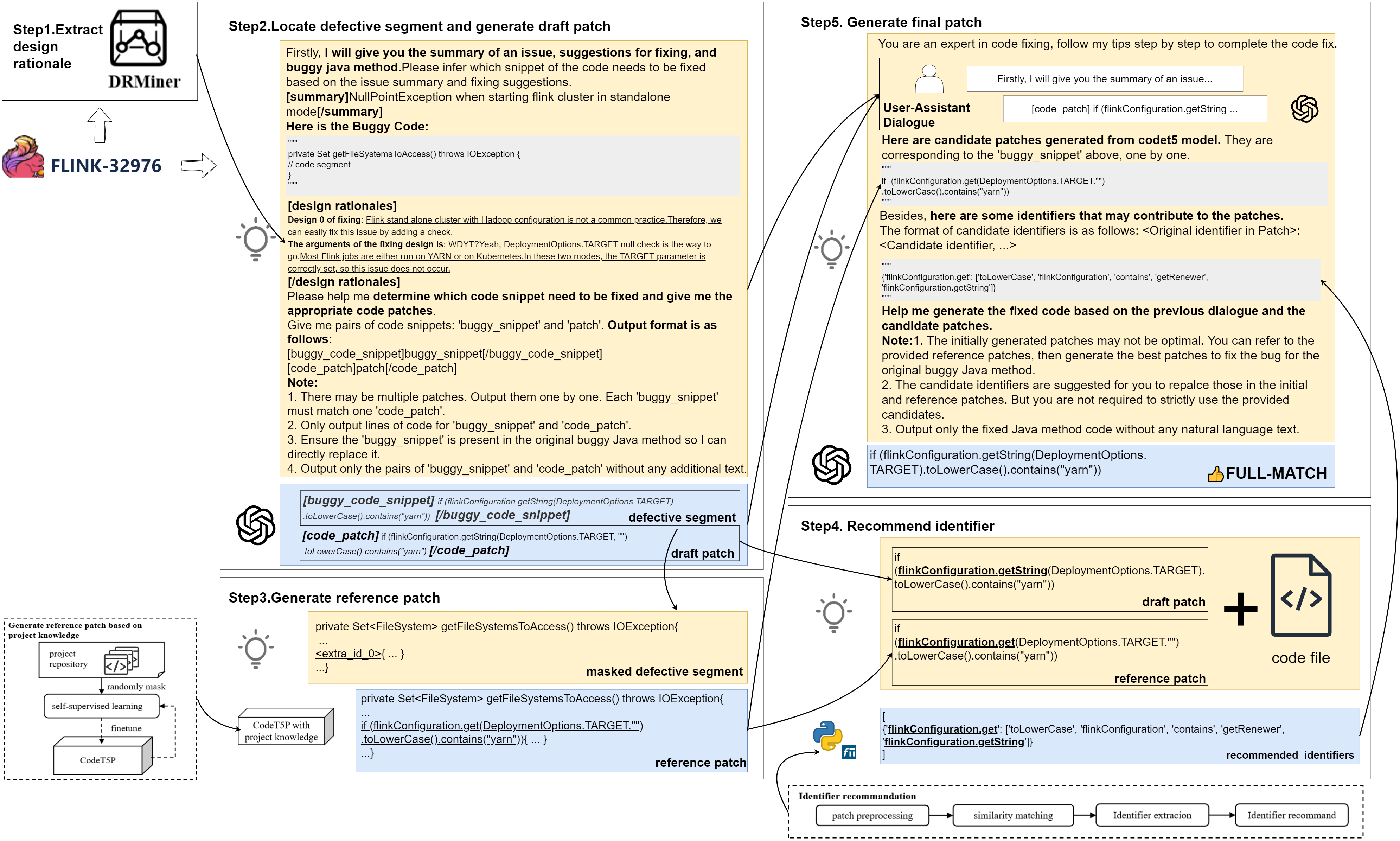}
        \caption{Overall Workflow of \textit{DRCodePilot} on FLINK-32976.}
        \Description{}
        \label{fig:OverallFrame}
    \end{figure*}

\subsection{Design Rationale Acquisition}
\label{subsec:DRAcquisition}
 The goal of this phase is to distill solutions and their respective arguments regarding specific issues, i.e., design rationales. Various methods have been proposed for extracting designs or solutions from diverse documents within the open-source community, such as mining issue-solution pairs from live chats \cite{shi2021ispy}, identifying question-answer dynamics \cite{2021Chatterjee_AutomaticExtractionofOpinion-BasedQA}, and uncovering design-centric discussions in pull requests \cite{viviani2021locating_LatentDesignInformationinDeveloperDiscussions}, among others. For our purposes, we adopt the \textit{DRMiner} approach \cite{zhao2024novel}, which specializes in deriving design rationales directly from issue logs using advanced LLMs. In essence, it combines tailored prompt instructions with specific heuristic features to pinpoint design-related text and align solution-argument pairs. Its superior performance has been validated across 2092 sentences from 30 issues spanning three publicly available open-source systems. 
   
    
\subsection{Defective Segment Location and Draft Patch Generation}
\label{subsec:draftGeneration}
       
    \textit{DRCodePilot} identifies the defective segment within the buggy function and generates the initial patch using the advanced LLM, GPT-4. GPT-4 is selected for its exceptional performance across a range of natural language tasks and for its capability to generate code based on provided instructions \cite{bubeck2023sparksgpt-4}.

    DRs extracted from issue logs with DRMiner are integrated into our prompt structure. As demonstrated in Figure \ref{fig:OverallFrame}, our prompt is composed of five sections:

    (1) \emph{An Instruction}: This is crafted to guide GPT-4 in identifying the defective segment and generating a patch based on the issue summary and fixing suggestions. 
    (2) \emph{Issue Summary}: Enclosed by \textit{[Summary]} tags, this part briefly outlines the problem.
    (3) \emph{Buggy Code}: A snippet of code, typically a function related to the issue.
    (4) \emph{Design Rationale}: These design rationales are drawn from the issue comments by the tool of DRMiner.
    (5) \emph{Output Instruction}: This specifies how the identified defective segments and the patches should be formatted.

    To ensure high-quality output, we impose four program-repair oriented guidelines for GPT-4 to adhere to rigorously:
    a) If there are multiple patches, they should be presented sequentially.
    b) The \textit{buggy\_snippet} and \textit{patch} must be expressed as lines of code, without any natural language text.
    c) It is imperative that the \textit{buggy\_snippet} is directly excerpted from the original buggy Java method.
    d) Outputs should be limited strictly to pairs of  \textit{buggy\_snippet} and \textit{patch}, without any supplementary natural language explanation.

\subsection{Reference Patch Generation}
\label{subsec:referenceGeneration}

    Advanced general-purpose LLMs are adept at generating code snippets from provided prompts, yet they may fall short in crafting precisely correct patches due to a lack of specific knowledge about the particular projects. To counter this, we utilize CodeT5P \cite{wang2023codet5p}, a smaller model amenable to fine-tuning within a project's context, to supply reference patches as feedback for GPT-4. This approach allows GPT-4 to reassess and refine its responses when necessary.

    Our methodology for generating reference patches is inspired by the principles set forth in \cite{xia2023_ThePlasticSurgeryHypothesis}. As depicted in Step 3 of Figure \ref{fig:OverallFrame}, during the reference phase, we obscure the fault-ridden segments previously identified by GPT-4 in Step 2 using the marker <extra\_id\_0>.  We selected CodeT5 due to its proficiency, having been pretrained on unimodal code corpora and bimodal code-text data, endowing it with considerable programming expertise \cite{xia2023_ThePlasticSurgeryHypothesis}. To tailor it to the coding conventions of the target project, we further fine-tuned it on the project's codebase. Additionally, given the substantial resources required to train larger models, we opted for the 220M variant of CodeT5P in our study.
    

    \begin{figure}[htb!]
        \centering
        \includegraphics[ trim = {0, 0, 0, 0}, clip, width=0.5\textwidth]{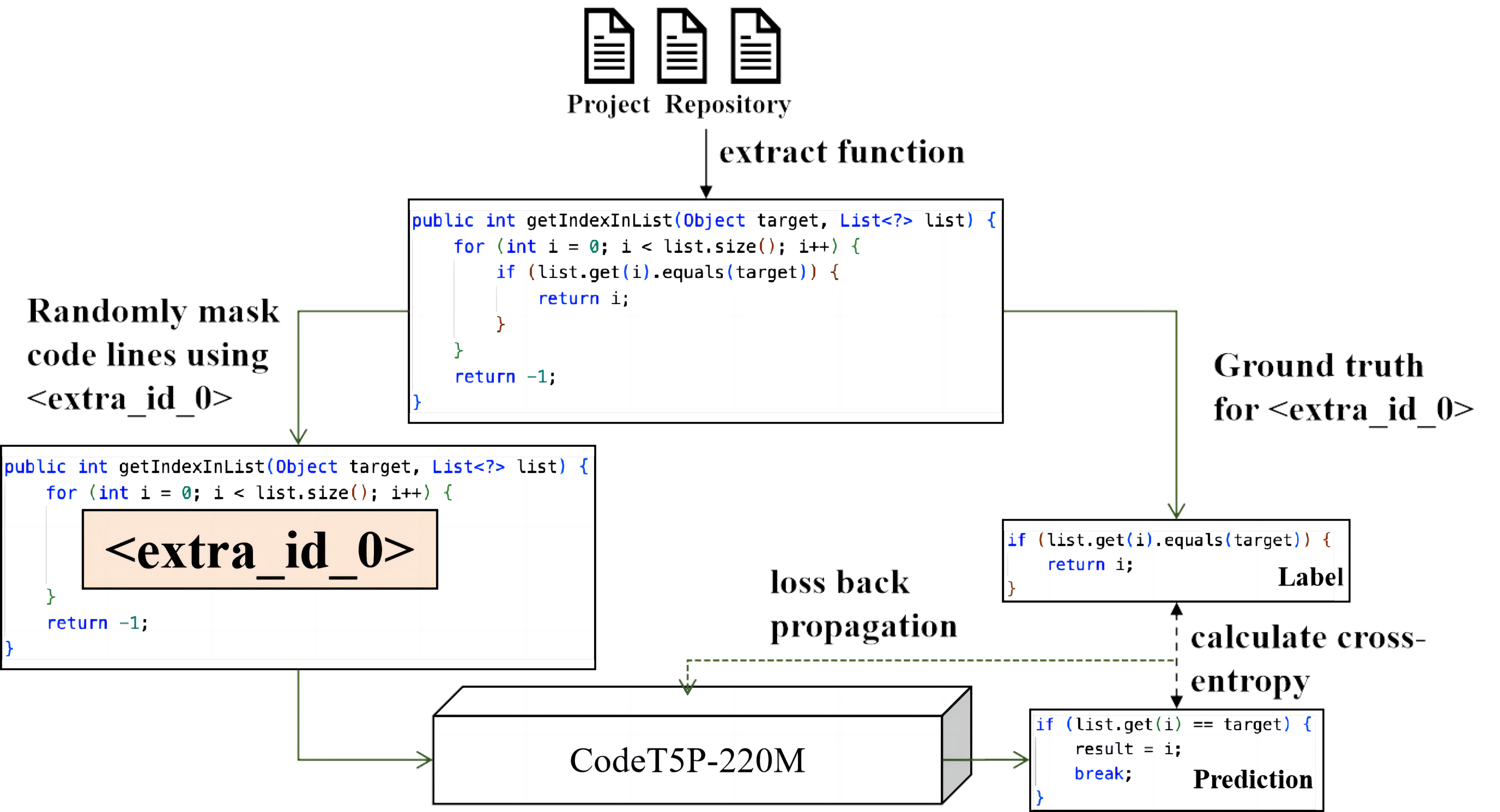}
        \caption{Fine-tuning Process of CodeT5P-220M for Project Knowledge}.
        \Description{}
        \label{fig:codeT2Ptrain}
    \end{figure}
    
    Fig. \ref{fig:codeT2Ptrain} shows the fine-tuning process. First, we build a training corpus from the project's codebase. For each Java file within this repository, we build an abstract syntax tree (AST). We then traverse through this AST to identify and extract the content of function nodes—termed ``methods'' in Java. To promote code conciseness, we exclude all comments and blank lines. Functions comprising less than three lines are also excluded, ensuring that only those with sufficient complexity are considered. The bodies of these methods are masked using \emph{<extra\_id\_0>}, aligning with CodeT5P’s pre-training protocols.

    To augment the fine-tuning corpus, we create multiple masked instances for each method so that every line within a method body gets masked at least once across different samples. Specifically, we designate the number of lines to be masked, denoted as $\alpha$, which ranges from one to five. This range is informed by one observation suggesting that most defects in maintenance rarely exceed five lines \cite{Dissection_of_a_bug_dataset_Sobreira}. Subsequently, we mask lines randomly based on the chosen scale $\alpha$. In subsequent iterations, we select another quantity, \emph{$\beta$} (ranging from 1-5, ensuring that $\alpha + \beta \leq n$, where \emph{n} represents the number of lines in the method body), and proceed to mask different lines based on this new count. This process is repeated until all lines have been subjected to masking.

Once the training corpus is prepared, we fine-tune the model with the loss function of cross-entropy. During this phase, CodeT5P absorbs project-specific knowledge, including method names and variable identifiers, crucial for rectifying issues. This learning is vital because many bugs can be remedied using code snippets from elsewhere in the source file or elsewhere within the project \cite{Barr14fse}.


\subsection{Identifier Recommendation} 
\label{subsec:identifierRecommendation}
   
   Draft patches and reference patches may contain identifiers that are semantically similar but incorrect in the project context. As illustrated in Fig. \ref{fig:OverallFrame}, the reference patch incorrectly uses the \emph{get()} call instead of the correct \emph{getString()}. At this stage, we aim to offer identifier recommendations in these two kinds of patches as additional feedback to GPT-4 for further refinement.
   
   Error-prone identifiers are detected based on a hypothesis: if an identifier appears solely in the generated patch and not elsewhere in the project code files, it is likely to be flawed. Considering the established observation that many solutions for bugs can be found within the same file \cite{Barr14fse}, we source recommended identifiers just from the Java file containing the defect.

   The core strategy for candidate identifier recommendation involves locating the most similar identifiers within code snippets that resemble the patch. Specifically, we divide the lengthy Java file into snippets matching the length of the given patch. We then calculate the CodeBLEU \cite{ren2020codebleu} score between each snippet and the patch, utilizing CodeBLEU due to its widespread application in measuring code similarity, which accounts for both semantic and structural similarities. From the three snippets that exhibit the highest similarity, we extract identifiers and evaluate the cosine similarity between the vectors of these candidate identifiers and the error-prone identifier found in the patch. The top-3 identifiers are preserved for consideration.


    Additionally, we compute the cosine similarity for identifiers outside the top-3 similar code snippets yet present in the defective Java file, selecting another set of top-3 identifiers as supplementary options. These may overlap with those chosen from the similar snippets. Consequently, for each error-prone identifier in the given patch, we provide up to six possible replacement candidates.

\subsection{Final Patch Generation}
    \label{subsec:finalPathGeneration}
    
    In the final step, GPT-4 adjusts its initial patches by incorporating feedback from Step 3 and Step 4. Particularly, the model's responses are reformulated into a User Assistant-dialogue format, complemented with the draft and reference patches along with suggested identifiers. Using this context, GPT-4 crafts the enhanced patch.

    This refinement occurs during Step 5, as illustrated in Fig. \ref{fig:OverallFrame}. Here, our prompt carefully defines GPT-4's role as a proficient code-fixing expert. By doing so, the model is directed to tap into its vast reservoir of training data, enabling the provision of sophisticated, expert-level code rectifications \cite{kong2024Role-Play}. Furthermore, it follows a deliberate ``follow my tips step by step'' chain-of-thought procedure \cite{wei2023chainofthought} (i.e., creating an initial patch, incorporating feedback, and refining the solution), which facilitates a clear understanding of the troubleshooting progression.

    Within the reference patch area, we supply exemplars of efficient solutions to steer GPT-4 toward compliance with established coding norms and best practices. GPT-4 can also discover and correct fine-grained errors more effectively based on identifiers recommendations, thereby improving the usability of the final generated patch. Additionally, we outline three distinct output directives to guarantee that the model's output aligns with the project's goals, ensuring the final code is not only structured but also easy to maintain.
\section{Experimental evaluation}
\label{sec:exp}

\subsection{Research Questions}
\label{subsec:RQs}

To evaluate the capabilities of \textit{DRCodePilot} in resolving real life software issues, we answer the following research questions:

\noindent\textbf{RQ1 (Baseline Comparison)}: To what extent can \textit{DRCodePilot} generate patches that effectively fix bugs?


\noindent\textbf{RQ2 (Ablation experiment)}: Are the injected design rationale and feedback-driven refinement effective in \textit{DRCodePilot}?


\noindent\textbf{RQ3 (DR with other LLMs)}: What is the effect of integrating the extracted DRs into other LLMs, and does the quality of DRs matter?

\subsection{Data Preparation}
\label{subsec:data}

Our experimental benchmark is composed of two critical elements: real-world issues enriched with pertinent design rationale discussions, and the corresponding human-crafted patches. While Defects4J \cite{Just2014Defects4J} holds the title as the most prevalent benchmark in the APR domain, it lacks incorporation of developers' discussions. SWE-bench \cite{jimenez2024swebench} has emerged as a benchmark that is rapidly gaining attention in the field. While it includes valuable developers' discussions, it unfortunately lacks critical meta-data labels such as the creator's name, comment participants, and comment indexing. This omission presents challenges for effectively extracting design rationales.

To address these deficiencies, we have developed our benchmark by selecting issues from the open-source systems Flink and Solr on Jira, ensuring comprehensive meta-data collection from issue logs, and correlating these with relevant GitHub pull requests (PRs) by tracing back through the provided issue links. Our focus on Flink and Solr stems from their status as emblematic large-scale software systems grappling with a wide array of problems.

We refine our dataset to only include instances where issues are closed and PRs are merged, which affords us the use of these human-written patches as verified ground truth. Harmonizing with Defects4J's approach, our filtration process scrutinizes patch files, guided by the parameters delineated in Table \ref{tab:DRCodePilotDataFilterCriteria}. Our priority lies with Java source file fixes that pivot on code logic—excluding unit tests, configuration files, and the like—and we narrow our scope to PRs altering solely one source file, with a maximum threshold set at 22 lines of change. This cut-off aligns with findings from Defects4J where 95\% of patches adjust under 22 lines \cite{Dissection_of_a_bug_dataset_Sobreira}.

The data presented in Table \ref{tab:patchcodelinechanges} showcases the distribution of `gold' patches based on the number of modified code lines within our benchmark. Notably, the majority of issues are resolved by altering fewer than 10 lines of code. This benchmark, inspired by Defects4J, includes issues that are, on average, less complex than those found in SWE-bench—judging by the scale of changes in lines and files. The rationale for this approach is to primarily investigate the extent to which design rationales embedded in developers' discussions can aid large language models in issue resolution. To this end, we have chosen to match the complexity level found in traditional Defect4J benchmarks. Looking ahead, we intend to tackle more challenging issues by incorporating design rationales from comments.

Furthermore, there is a noticeable disparity in data volume between Solr (1.1k stars) and Flink (23.3k stars), which likely stems from the different levels of popularity the two projects enjoy on GitHub; lower popularity often results in fewer commits, issue reports, and subsequent fixes.


It is also important to highlight that we did not build an execution-based evaluation pipeline for these patches. Although both projects feature comprehensive unit test suites, the immense evolution of these projects over time—Flink over 10 years and 20 years for Solr—has led to significant changes in both Maven project structures and their library dependencies. This evolution posed challenges in compiling test suites for an execution-based evaluation pipeline, as outdated library versions caused unresolved dependencies. Consequently, even among expert-written and merged code patches, less than 1\% are amenable to executable evaluations.

Additionally,  to avoid unfair comparisons caused by patches appearing directly in comments, all code snippets in comments were replaced with [code] tags.

\begin{table}[!htbp]
\caption{Key Indicators of Our Benchmark. }
\centering
\small
    \begin{tabular}{p{4.2cm}|p{3.5cm}}
    \hline
    \textbf{Indicator} & \textbf{Condition} \\
    \hline
    Included projects & Solr and Flink \\
    \hline
    Data collection source &  Closed issues and merged PRs \\ 
    \hline
    Issue metadata &  Yes \\ 
    \hline
    File types for fixes & Java source files \\
    \hline
    Number of files modified per PR & One source file \\
    \hline
    Number of code lines changed per PR & Maximum of 22 lines \\
    \hline
    Length of issue comments & 	Unlimited \\
    \hline
    \end{tabular}
\label{tab:DRCodePilotDataFilterCriteria}
\end{table}

\begin{table*}[!htbp]
    \caption{Distribution of the Patches over the Changed Code of Lines in Our Benchmark.}
    \centering
    \begin{tabular}{l|c|c|c|c|c|c}
        \hline
        & $1 \leq n < 5$ & $5 \leq n < 10$ & $10 \leq n < 15$ & $15 \leq n < 20$ & $20 \leq n < 22$ & Total \\
        \hline
        Flink & 514 (71.99\%) & 139 (19.47\%) & 40 (5.60\%) & 12 (1.60\%) & 9 (1.26\%) & 714 \\
        \hline
        Solr  & 125 (55.80\%) & 75 (33.48\%)  & 12 (5.36\%) & 4 (1.79\%)  & 8 (3.57\%) & 224 \\
        \hline
        \end{tabular}
    \label{tab:patchcodelinechanges}
\end{table*}

\subsection{Experiment Setting}

    
    \noindent \textbf{Evaluation Metrics}. We employ two metrics to assess the effectiveness of the various APR approaches involved. First, we tally the number of \emph{full-match} patches—those that are identical to the provided gold patches. We did not utilize the widely applied pass@k criterion based on test cases for two primary reasons. Firstly, as outlined in Section \ref{subsec:data}, constructing test cases for legacy Java projects presents significant challenges. More importantly, this choice is predicated on the reality that in professional settings, patches must undergo a rigorous manual review by multiple engineers before they can be deemed suitable for integration into the project—even if they succeed in passing predefined tests.
    Hence, we employ the `Full-Match' criterion to select patches that are identical to the manual patches and ready for use as solutions.
    
    Second, we apply \emph{CodeBLEU} \cite{ren2020codebleu}, a widely recognized metric for evaluating code generation quality that compares the semantic and syntactic precision of the generated code against gold code. It should be noted that CodeBLEU is better suited for assessing the quality of code at the level of functions or code snippets, due to its comprehensive analysis of components like syntactic accuracy, data flow, and logical structure, which typically span multiple lines of code. As per Table \ref{tab:patchcodelinechanges},  it is observed that over half of the patches involve fewer than five lines of change. Therefore, we calculate CodeBLEU scores for the entire repaired function, meaning the original function wherein the flawed segment has been modified with the suggested patch.

    \noindent \textbf{Baselines}. The baseline methods selected in this experiment include four state-of-the-art large code models: 
    \begin{itemize}[leftmargin=3mm]
        \item  \emph{StarCoder2} \cite{lozhkov2024_starcoder2}: Through a new training technique called CodeBERTa, StarCoder2 improves the model's generalization ability and understanding of programming languages without increasing the number of parameters. The model version used in this experiment is StarCoder2-15B.
        \item  \emph{CodeLlama} \cite{roziere2023codellama}: Based on pre-training from LLaMA2 \cite{touvron2023llama2}, CodeLlama enhances coding capabilities, better follows human instructions, and understands zero-shot tasks. The model version taken for this experiment is CodeLlama-7B.
        \item \emph{CodeShell} \cite{xie2024codeshell}: Using high-quality pre-training data, CodeShell fuses the core features of StarCoder \cite{li2023_starcoder} and LLaMA2, supports code-specific generation methods, and has a high-performance and easily extensible context window architecture. CodeShell has a model parameter count of 7B. It outperforms models with the same number of parameters on the HumanEval test.
        \item \emph{GPT-3.5 (gpt-3.5-turbo-1106)}: We choose GPT-3.5 because it is free to use and is widely used as an effective aid to coding, and researchers can easily access and utilize it for experimental comparisons.
        \item \emph{GPT-4 (gpt-4-1106-preview)} : Successor to GPT-3.5, but with major enhancements in model size, training data, and capabilities, reportedly up to 1 trillion parameters\footnote{https://the-decoder.com/gpt-4-has-a-trillion-parameters/}. It is often considered to represent the highest level of the current large language models. 
    \end{itemize}

\subsection{Baseline Comparison (RQ1)}
    
    Table \ref{tab:comparison} showcases the comparative evaluation of our design rationale-driven approach, \textit{DRCodePilot}, against baselines that generate patches directly from buggy functions.  We detail the number of full matches and their proportion relative to all samples. Our method outperforms the baselines across both systems in the benchmark, as indicated by the number of full-match instances and CodeBLEU scores. Among the baselines, GPT-4 leads, followed by GPT-3.5, CodeLlama, StarCoder2, and CodeShell.

    \textit{DRCodePilot} significantly surpasses the best-performing baseline, GPT-4, in generating full matches. For instance, it achieves 109 full matches for Flink, which is about 4.7 times more effective than GPT-4. In Solr, the number of full matches by \textit{DRCodePilot} is up to 3.6 times that achieved by GPT-4. The CodeBLEU scores also position \textit{DRCodePilot} at the forefront within both systems' benchmarks. The increase in CodeBLEU, while not as pronounced as the full-match improvement, can be attributed to the need for calculating CodeBLEU over complete functions \cite{ren2020codebleu}, where the altered code may only constitute a minor segment of the whole function. This implies that generated patches have a constrained influence on CodeBLEU scores. Nonetheless, this metric still reflects the similarity between the patches and the 'gold' standard answers.

    Upon cross-examining the two benchmarks, we noted that the full-match ratio of \textit{DRCodePilot} in Flink (15.27\%) exceeds that in Solr (8.03\%), a trend consistent among the other baselines. Excluding \textit{DRCodePilot} and GPT-4, all other baselines register better CodeBLEU scores in Flink compared to Solr. A contributing factor is the greater complexity of issues in Solr, as indicated by a higher proportion of patches with changed lines in the ranges of [5,10),[15,20), [20,22), and much less proportion in the range of [1,5), compared to those in Flink (refer to Table \ref{tab:patchcodelinechanges}).

    \begin{table*}[!htbp]
        \centering
    \caption{Comparing the Performance of Different Methods on FLink and Solr }
        \label{tab:comparison}
        \scalebox{1.0}{
            \begin{tabular}{c|cc|cc}
                \toprule
                \multirow{2}{*}{\textbf{Approach}} & \multicolumn{2}{c|}{\textbf{Flink} (Total: 714 samples)} & \multicolumn{2}{c}{\textbf{Solr} (Total: 224 samples)} \\
                \cmidrule(lr){2-3} \cmidrule(lr){4-5}
                & Full-Match & CodeBLEU & Full-Match & CodeBLEU  \\
                \midrule
                CodeShell & 6(0.84\%) & 0.71  & 0 & 0.61  \\
                CodeLlama & 8(1.12\%) & 0.64 & 2(0.89\%) & 0.62 \\
                StarCoder2 & 9(1.26\%) & 0.74  & 0 & 0.64  \\
                GPT-3.5 & 15(2.10\%) & 0.72  & 1(0.45\%) & 0.69  \\
                GPT-4 & 23(3.22\%) & 0.74  & 5(2.23\%) & 0.77  \\
                \textbf{DRCodePilot} & \textbf{109}(15.27\%) & \textbf{0.78}  & \textbf{18}(8.03\%) & \textbf{0.80}  \\
                \bottomrule
            \end{tabular}
        }
    \end{table*}

\subsection{Ablation Study (RQ2)}

    Three types of knowledge play a crucial role in our \textit{DRCodePilot}: design rationale for initial patch generation, reference patches and identifier recommendations for patch optimization. We aim to assess the contribution of each knowledge type to the quality of the final patch produced.

    To accomplish this, we crafted three variants of \textit{DRCodePilot} for our study: 1) \emph{DRCodePilot-DR}, which omits design rationale from the prompt during the initial patch creation phase (removing Step 1 and changing Step 2 in Fig.\ref{fig:OverallFrame}). 2) \emph{DRCodePilot-PF}, which does not incorporate the generation of reference patches (removing Step 3 from DRCodePilot shown in Fig. \ref{fig:OverallFrame}). And 3) \emph{DRCodePilot-ID}, which removes identifier recommendations from the process (excluding Step 4 from \textit{DRCodePilot} illustrated in Fig. \ref{fig:OverallFrame}). All other experimental variables and procedures are kept consistent across the three variants. 

    These three variants were assessed using our benchmark, with the final results presented in Table \ref{tab:DRCodePilot_xiaorong}. To better illustrate the comparative performance, we calculate the performance drop ratio for each variant relative to the original \textit{DRCodePilot} across the two metrics. Specifically, if the metric value for \textit{DRCodePilot} is denoted as $\alpha$, and the corresponding value for one variant is $\beta$, then the performance drop ratio is computed as $\frac{\alpha - \beta}{\alpha}$. A positive ratio, indicated by a downward arrow and red font, signifies that excluding a certain type of knowledge has weakened the performance of that variant.

    \begin{table*}[!htbp]
    \centering
    \caption{Ablation Results on Our Benchmark.}
    \label{tab:DRCodePilot_xiaorong}
        \begin{tabular}{c|cc|cc}
            \toprule
            \multirow{2}{*}{\textbf{Approach}} & \multicolumn{2}{c|}{\textbf{Flink}} & \multicolumn{2}{c}{\textbf{Solr}} \\
            \cmidrule(lr){2-3} \cmidrule(lr){4-5}
             & Full-Match & CodeBLEU & Full-Match & CodeBLEU \\
            \midrule
            DRCodePilot-DR & 20 \textcolor{red}{($\downarrow81.65\%$)} & 0.713 \textcolor{red}{($\downarrow9.17\%$)} & 1 \textcolor{red}{($\downarrow94.44\%$)} & 0.762 \textcolor{red}{($\downarrow4.87\%$)} \\
            DRCodePilot-PF & 109 (-) & 0.783 \textcolor{red}{($\downarrow0.26\%$)} & 19 \textcolor{black}{($\uparrow5.55\%$)} & 0.800 \textcolor{red}{($\downarrow0.12\%$)} \\
            DRCodePilot-ID & 105 \textcolor{red}{($\downarrow3.67\%$)} & 0.781 \textcolor{red}{($\downarrow0.51\%$)} & 17 \textcolor{red}{($\downarrow5.55\%$)} & 0.794 \textcolor{red}{($\downarrow0.87\%$)} \\
            DRCodePilot & \textbf{109} & \textbf{0.785}  & \textbf{18} & \textbf{0.801}  \\
            \bottomrule
        \end{tabular}
    \end{table*}

    Generally, the absence of any knowledge type tends to diminish the model's performance, affecting both the count of full-match patches and CodeBLEU scores. Notably, the most pronounced performance decrease is observed with \textit{DRCodePilot-DR} in terms of both the number of full matches and CodeBLEU metrics. The count of full-match patches plummeted by 81.65\% and 94.44\%, while the CodeBLEU scores fell by 9.17\% and 4.87\% across the two systems, respectively. This suggests that incorporating design rationales into the initial patch generation prompts can significantly improve patch quality, as they offer pertinent discussions that catalyze GPT-4's issue resolution capabilities. It's unsurprising that the decline in CodeBLEU is more modest than in full-match counts due to the relatively small proportion of changed lines within the entire defective functions.

    Additionally, the greater reduction in full-match patches observed in Solr compared to Flink may suggest that design rationales assume a more crucial role in resolving complex issues, as evidenced by the larger amount of code line changes in Solr relative to Flink. We aim to further investigate this hypothesis in future work.

    The second most noticeable performance decline is seen with \textit{DRCodePilot-ID}, that is, upon removing identifier recommendations (Step 4), with drops of 3.67\% and 5.55\% in full-match counts, and 0.51\% and 0.87\% in CodeBLEU, respectively, for the two systems. These figures reveal that even an advanced code-focused LLM like GPT-4 can err in generating precise correct identifiers during issue repairs. However, our feedback-based, self-reflective prompt guides it towards re-evaluation and refinement of unsuitable identifiers.

\begin{figure}[htb!]
        \centering
        \includegraphics[ trim = {0, 0, 0, 2cm}, clip, width=0.45\textwidth]{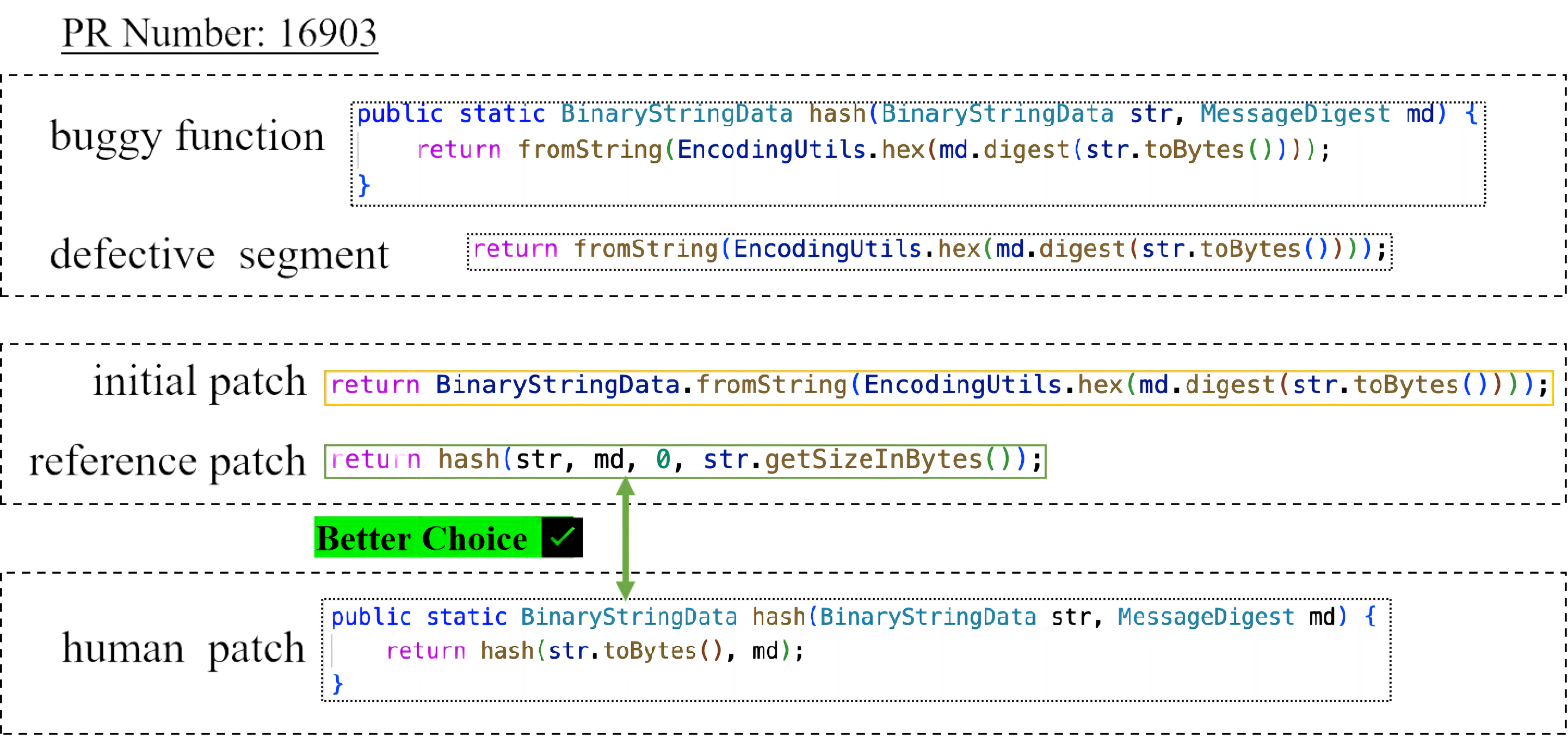}
        \caption{One Example Indicating the Usefulness of Project Knowledge-based Reference Patch}.
        \Description{}
        \label{fig:projectknowledge}
    \end{figure}

    Upon reviewing the outcomes of \textit{DRCodePilot-PF}, we observe that the count of exact-match patches remains constant in Flink and experiences a marginal increase in Solr. Concurrently, all CodeBLEU scores within these two systems exhibit a downturn. This indicates that GPT-4 does not merely replicate the reference patch but rather assimilates pertinent information from such feedback to refine its responses. Furthermore, we aim to investigate scenarios where project-specific reference answers contribute to the creation of superior quality patches. A thorough examination of the exact-match patches produced by both \textit{DRCodePilot} and \textit{DRCodePilot-PF} suggests that when design rationales behind issue resolutions are discernible, the sophisticated GPT-4 model demonstrates proficiency in generating high-caliber patches. In contrast, reference patches derived from simpler models can erode its confidence, leading to the displacement of exact matches. On the flip side, in cases where commentary on issues is scant, reference patches seem advantageous. To elucidate this phenomenon, we provide a representative example in Figure \ref{fig:projectknowledge}.

    \emph{One example illustrating the utility of project knowledge-based reference patches:} Take the issue FLINK-23579\footnote{\url{https://issues.apache.org/jira/browse/FLINK-23579}} as an instance. The associated issue log includes a mere pair of comments, with the first pinpointing the root cause yet neither suggesting a fix. Consequently, we cannot incorporate any design rationale into the initial patch generation prompt. GPT-4 formulates the initial patch solely based on internal knowledge, contrasting with the reference patch crafted using project-specific insights. The reference patch highlights the overloaded \emph{hash()} function, which aligns with the actual fix implemented. Evidently, the reference patch closely matches the engineer's approach. This exemplifies how project knowledge can effectively supplement design rationales, especially when explicit solutions are absent, and the model must discern the most project-appropriate solution among several possibilities.

    \subsection{General Impact of Design Rationale (RQ3)}

    In this section, we aim to address three sub-questions. Firstly, our framework has thus far only utilized design rationales in conjunction with GPT-4. We are intrigued by the broader applicability of this approach in APR tasks. Specifically, we seek to understand \emph{the extent to which these design rationales may bolster the effectiveness of different models engaged in program repair \textbf{(RQ3.1)}}.
      
    Secondly, the design rationales we have employed were autonomously derived using the DRMiner methodology \cite{zhao2024novel}. However, the evaluation outcomes presented in the originating paper indicate that the extraction process is not infallible, evidenced by inaccuracies and omissions. In light of this, our experiment sets out to ascertain \emph{the degree of improvement in program repair performance achievable through the integration of meticulously hand-annotated design rationales \textbf{(RQ3.2)}}.

    Moreover, given that the design rationales are gleaned from developers' comments, a natural question arises: \emph{can advanced general-purpose LLMs independently comprehend these comments and distill pertinent insights to facilitate issue resolution \textbf{(RQ3.3)?}}

     To explore the above sub-questions, we selected three baseline models capable of interactive prompting: CodeLlama, GPT-3.5, and GPT-4. 
     

    
    Since expert annotations are involving, we randomly selected 61 issues that featured extensive comments from our benchmark. We focus on the quantity of comments because design rationales are often given in the comments, and we aim to assess the impact of design rationale quality. In other words, the presence of design rationales is a necessary condition for our issue selection.
    
    We enlisted 16 participants with varied backgrounds, including two professors with over 12 years of academic experience in computer science; seven Ph.D. students and six postgraduate students engaged in both academic and industrial software research; along with one software engineer boasting more than six years of development experience at internet companies.
    
    The annotation process involved independent analyses and joint discussions among three participants. Specifically, each issue was assigned to two of them, who would annotate the design rationales independently. Once the independent annotations were completed, a discussion involving the two initial annotators and a third participant took place. In cases of disagreement, a majority vote decided the final labeling outcome.
    
    The evaluation results are summarized in Table \ref{tab:ManualLabeling}, from which we can draw the following insights:
    \begin{itemize}[leftmargin = 3mm]
        \item \emph{Generality (RQ3.1)}: Our approach and all three baseline models demonstrate promising results when utilizing design rationales, even when these are imperfectly extracted by a tool (+DR). Without design rationale support, the three baseline models could only generate one or two exact match patch corresponding to the reference solution. However, when provided with design rationales—even noisy ones—the figures rise to 11 and, impressively, 19 for the more advanced GPT-4 (11x-19x).

        \item \emph{Better Design Rationale, Better Patch (RQ3.2)}: Across the board, more full-match patches are produced by all baselines and our \textit{DRCodePilot} with the deliberately annotated design rationales (+DDR). Nevertheless, it's noteworthy that the CodeBLEU score for GPT-3.5's output experienced a decrease. A probable explanation is that \emph{DRMiner} favors recall over precision \cite{zhao2024novel}, indicating that the extraction process includes some sentences unrelated to design rationale. However, GPT-3.5 can find useful hints from these noisy information to enhance program repair.

        This reveals the disparity between automated tools for design rationale extraction and expert annotations. Future research is expected to narrow this gap and enhance the efficiency of utilizing design rationales. In conclusion, the results confirm the influence of design rationales on repair results, with more precise design rationales leading to better code repair outcomes.
        
        \item \emph{Enhanced Performance with Developer Comments Over No Comments, Yet Surpassed by Design Rationale Integration (RQ3.3)}: The inclusion of original developer comments (+AC) distinctly enhances program repair performance for all models compared to those that do not incorporate discussions. An evident increase in the number of full-match patches is observed across all four models, supporting the findings of Panthaplackel et al. \cite{DBLP:conf/emnlp/PanthaplackelGL22} which emphasize the constructive role of developer conversations in bug-fixing activities.

        Furthermore, the degree of this enhancement varies among models of different complexities. CodeLlama and GPT-3.5 demonstrate around an elevenfold increase in full-match patches, while GPT-4 shows a twelfthfold improvement. Our \textit{DRCodePilot} especially stands out, achieving twenty-two times more full-match patches, showcasing the considerable advantage of directly leveraging developer comments. The rationale is straightforward: advanced models such as GPT-4 are capable of assimilating and deciphering valuable solution-oriented design information from developer comments.

        However, upon examining the data for +AC, +DR, or +DDR, it becomes clear that models utilizing design rationales consistently outperform those relying solely on original comments, in terms of both full-match patches and CodeBLEU scores. This suggests that despite the presence of design rationales within the comments, general models struggle to pinpoint these key insights amidst the intertwined discussions, and this challenge persists even for sophisticated models like GPT-4. Therefore, there is a need to employ an additional tool to initially extract the solution-oriented design information before incorporating it into prompts to guide GPT-4 towards generating superior patches.
        
        \end{itemize}

        Additionally, it is evident that our \textit{DRCodePilot} retains its frontrunner status than the baselines, even with the baseline models showing enhanced performances through the incorporation of design rationales. Nevertheless, the performance gap between GPT-4 and our \textit{DRCodePilot} is relatively small compared to other baselines. As previously mentioned, the 61 issue logs contain abundant comments, which suggests the presence of extensive discussion and high-quality design rationale. In such scenarios, GPT-4 is capable of generating high-quality patches. However, we incorporate feedback mechanisms and self-reflective prompts, recognizing that it's not always possible to guarantee the existence of high-quality comments for all issues, particularly new and unresolved ones.


\begin{table}[!htbp]
    \centering
    \caption{The Effect of Design Rationale Quality on Varied Models' Performance (with 61 cases). ``+DR'' denotes the inclusion of automatically mined design rationales, ``+DDR'' signifies the addition of deliberately manual-annotated design rationales, and ``+AC'' denotes the incorporation of all developer comments in each issue log.}
    \label{tab:ManualLabeling}
        \begin{tabular}{l|c|c}
            \hline
            \textbf{Approach} & \textbf{Full-Match}  & \textbf{CodeBLEU} \\
            \hline
            CodeLlama & 1 &  0.68 \\
            CodeLlama+DR & 11 &  0.75\\
            CodeLlama+DDR & 14 &  0.83 \\ 
            CodeLlama+AC & 11 & 0.80 \\ \hline
            
            GPT-3.5 & 1 &  0.75 \\
            GPT-3.5+DR & 11 &  0.71 \\
            GPT-3.5+DDR & 14 &  0.66 \\ 
            GPT-3.5+AC & 11 & 0.70 \\  \hline
            
            GPT-4 & 1 &  0.75 \\
            GPT-4+DR & 19 &  0.85 \\
            GPT-4+DDR & 21 &  0.84 \\ 
            GPT-4+AC & 19 & 0.87  \\
            \hline
            
            DRCodePilot w/o DR & 1 &  0.76 \\
            DRCodePilot (+DR) & 26 &  0.88 \\
            DRCodePilot+DDR & \textbf{26} & \textbf{0.88} \\
            DRCodePilot+AC & 22 & 0.86 \\
            \hline
        \end{tabular}
\end{table}

\label{subsec:eva}
\section{Related works}
\label{sec:relatedWork}

\subsection{Design Rationale Extraction}

The design rationale encompasses the alternative design options considered and decided upon throughout the software life cycle \cite{2001MurphySoftwarereflexion,Design_Knowledge_and_Design_Rationale:A_Framework_for_Representation_Capture_and_Use_ThomasR,Bridging_the_Gap_Between_Software_Architecture_Rationale}, as well as the justifications for accepting or rejecting certain alternatives. It is widely applied to assist human engineers in development and design activities \cite{A_Tool_for_Managing_Software_Architecture_Knowledge_Babar,Using_Rationale_to_Support_wang_978-1-60558-967-1,A_Web-based_Tool_for_Managing_Architectural_Design_Decisions_Capilla_0163-5948,ARationale-basedArchitectureModel_Tang_0164-1212,JANSEN2008536Documenting_afterthefact,ZHANG201313semanticrepresentation}. Effectively documenting design rationale is critical to maintaining the long-term health and vitality of a project, allowing software developers to understand past decisions and continue to implement these core design ideas in future updates and maintenance. 

In recent years, automatically mining design knowledge from various sources (such as emails, meeting notes, or open-source communities) has attracted widespread research interest \cite{shi2021first_LookatDevelopers,shi2021ispy,viviani2021locating_LatentDesignInformationinDeveloperDiscussions,srivastava2022argument_miningusingbertandself-attentionbasedembeddings,2021PankajeshwaraExtracting_RationaleforOpenSourceSoftwareDevelopment,AnalysisandDetectionofInformationTypesofOpenSourceSoftwareIssueDiscussions,panthaplackel-etal-2022-learning,Krasniqi2021Bug-fixing-Comments}, as a substantial number of decisions actually occur in informal discussions. For instance, Panthaplackel et al. \cite{panthaplackel-etal-2022-learning} generated descriptions of solutions by synthesizing relevant content. However, human evaluators scored the informativeness of the descriptions (using a scale of 1-5) at an average of 3.3, suggesting that the descriptions include useful information but does not capture the solution well.

A major challenge of this task is that these discussion texts are interleaved, noisy, and the expressions of design rationale are diverse and abstract \cite{AnalysisandDetectionofInformationTypesofOpenSourceSoftwareIssueDiscussions,CHANDRASEGARAN2013204knowledgeRepresentation}, which makes previous methods ineffective in acquiring design rationales from issue logs. The recently emerged DRMiner \cite{zhao2024novel}, leveraging customized features and large language models, achieves promising performance on the design rationale extraction. It effectively identifies solutions and their associated arguments from issue logs on Jira, laying a solid foundation for leveraging design rationale in software activity.

\subsection{Automatic Program Repair}

Automated program repair (APR) technologies have exerted significant influence across various domains such as software engineering, system security, and artificial intelligence \cite{A_Survey_on_Automated_Program_Repair_Techniques_Kai2023}. APR can broadly be classified into four categories: (1) \emph{Semantic search-based methods} \cite{GenProg_a_Generic_Method_for_Automatic_Software_Repair_Le2012,Shaping_Program_Repair_Space_with_Existing_Patches_and_Similar_Code_Jiang2018} analyze the structure and contextual information of code to find and recommend repair solutions. (2) \emph{Semantic constraints-based methods} guide the repair process by developing a set of constraint specifications, transforming program repair problems into constraint solving problems \cite{nopol2016xuan,On_the_efficiency_of_test_suite_based_program_repair_liu2020}. (3) \emph{Pattern-based techniques} reduce search space through template matching, which can be manually extracted or automatically mined to guide specific types of defect repairs \cite{kim2013_Automatic_patch-generation_learned_from_human-written_patches,le2016_historydrivenprogramrepair,koyuncu2019ifixr_Bugreportdrivenprogramrepair,liu2019tbar}. (4) \emph{Learning-based APR} learn the experiential knowledge of program repair from a large number of repair samples, having the flexibility and scalability to follow complex instructions and handle types of defects \cite{marquez2018empirical_studyofscalabilityframeworksinopensourcemicroservices-basedsystems,white2019sorting_transforming_programrepairingredients,li2020dlfix_Context-basedcodetransformation,lutellier2020coconut_combiningcontext-awareneuraltranslation,xia2022less_training_more_repairing_pleas,xia2023_ThePlasticSurgeryHypothesis,panthaplackel-etal-2022-learning}. Panthaplackel et al. obtain useful information from discussions (e.g. solution description) to enhance learning-based APR \cite{DBLP:conf/emnlp/PanthaplackelGL22}. They fine-tuned a sequence-to-sequence model to generate the fixed code given varying input context representations. 

Recent work has explored applying code large language models for APR \cite{li2023_starcoder,lozhkov2024_starcoder2,roziere2023codellama,xie2024codeshell,wang2023codet5p}. A prevalent paradigm involves formulating APR as a code generation task, and leveraging prompt engineering techniques to steer the model towards generating more effective patches\cite{prenner2022can_OpenAI,xia2023_automated_program_repair_in_the_era_of_large_pre-trained,xia2023chat-repair}. One challenge of leveraging LLM is to overcome code hallucination , where LLM sometimes generate code that appears plausible but fails to meet the expected requirements or executes incorrectly \cite{tian2024codehalu}. Some studies suggest fine-tuning the LLM for program repair to enhance its effectiveness \cite{jiang2023impact,silva2023repairllama,hossain2024Toggle}, but the cost is high. Other methods include providing feedback to the model based on external tools and guiding it to self-correct \cite{ji2023towards_Mitigating_Hallucination,shinn2023reflexion}, or providing useful information to the model such as bug reports \cite{motwani2023better,koyuncu2019ifixr_Bugreportdrivenprogramrepair}, retrieved API context \cite{zhang2024autocoderover}, etc. 

In this paper, we extract comprehensive design rationales (e.g. alternative solutions and arguments) , to guide LLM in generating patches using zero-shot inference. To better generate feasible patches, we further devise a feedback mechanism by addressing potential technical information gaps in design rationales.


\section{Discussion}
\label{discuss}

\subsection{Failure Case Analysis}
To better understand the conditions that influence DRCodePilot's performance, we analyzed 72 patches: 36 full-match patches (18 randomly selected from Flink and all 18 full-match patches from Solr) and 36 lower-quality patches (randomly selected in the same way, all with a CodeBLEU score below 0.5). Through comparisons, we find two scenarios that our \textit{DRCodePilot} may not perform as expected.

\begin{itemize}[leftmargin = 3mm]
     \item \emph{Quality and Richness of the Extracted DR largely impact the performance of \textit{DRCodePilot}}: The efficacy of \textit{DRCodePilot} is intrinsically linked to the quality and comprehensiveness of the extracted DRs. Challenges arise when DR is limited or absent—such as when developers' discussions go unrecorded (e.g., FLINK-28513)—resulting in subpar DRs that hinder DRCodePilot's performance. For simple issues, \textit{DRCodePilot} can generate fine patches. However, it stuggles with complex issues.
    
    \item \emph{Insufficient Contextual Knowledge in DRs}: Our approach hinged exclusively on the ``one-stop'' DR, which is directly gleaned from developer discussions. We refrained from incorporating supplementary context to fill gaps essential for a comprehensive understanding. The absence of such extended contextual knowledge can render the extracted DRs obscure, hampering DRCodePilot's capacity to infer solutions accurately, and consequently resulting in less-than-optimal results. We recognize three particular scenarios illustrative of this limitation:

\begin{itemize}[leftmargin=3mm]
    \item \textbf{Code References in DR Solutions:} If a DR refers to code specifics but omits the actual code snippet, like in SOLR-15896 where it instructs to `Just copy the \emph{toString} method and change the broken parts,' without including the \emph{toString} content, \textit{DRCodePilot} finds it challenging to generate patches aligned with manually crafted ones.

    \item \textbf{Involvement of External Links:} Instances where DRs point to external resources or related issues can deprive \textit{DRCodePilot} of critical context. An example is FLINK-28488, where a DR references FLINK-27487 for key details. However, our methodology is not designed to extract these particulars, thereby omitting the integration of vital insights.
    
    \item \textbf{Unclear Code Intent:} At times, developer discussions may highlight errors without specifying the correct behavior. This impedes DRCodePilot's patch quality, as in FLINK-15386, where a DR identifies a logical mistake but fails to convey the desired logic, leaving the model at odds with deriving an accurate fix despite recognizing the error.
\end{itemize}
\end{itemize}

\subsection{Threats to Validity}
The principal threat to the \emph{construct validity} in our study is associated with the choice of benchmark. Publicly available benchmarks, such as  Defects4J \cite{Just2014Defects4J} and SWE-bench \cite{jimenez2024swebench}, were not employed due to their lack of developer discussions or the meta-data labels of discussions —data critical for extracting design rationales. Accordingly, we constructed a new benchmark by aligning issues from Jira with their corresponding patches in GitHub. We believe that our custom benchmark will unlock novel possibilities for advancing design-guided program repair research.

We limited our evaluation to design rationales extracted solely from Jira, a single issue tracking system. Other platforms such as GitHub issues also host design discussions, which could affect the \emph{external validity} of our findings.  Given the analogous metadata structures between Jira and GitHub issues, we anticipate that our approach would be compatible. Nonetheless, further assessment across different platforms is warranted to validate this assumption.

Manual annotation often poses a threat to \emph{internal validity} due to potential inconsistencies or biases. To mitigate such risks, we employed standard procedures involving independent annotations followed by collaborative discussions for consensus building, ultimately deciding outcomes through majority vote.

Another potential threat to \emph{internal validity} stems from the content within developer comments. It is unsurprising that comments, or the design rationales derived from them, play a significant role if they contain explicit target code snippets. In practice, developer discussions are typically high-level and advisory in nature. Nonetheless, to minimize the impact of such rare occurrences, we ensure all code within the comments is obscured using a [code] token.
    
\subsection{Limitations}

\begin{itemize}[leftmargin = 3mm]
    \item \emph{Absence of Repair-Specific Baselines}: The baselines utilized in this study are general-purpose code LLMs. Specific bug-fixing models such as RepairLLaMA \cite{repairllama2023} were not selected because they do not fit our experimental context. Most specialized models require additional inputs, like the defective locations in the buggy function \cite{repairllama2023} or a buggy template \cite{liu2019tbar}. In our study, such supplementary information is not available.

    \item \emph{Evaluation Limited to Simpler Issues}: Echoing the benchmark of Defects4J \cite{Just2014Defects4J}, where 95\% of patches involve fewer than 22 lines of changed code, our dataset selection was similarly scoped to primarily assess the impact of design rationales on APR with relatively simpler issues. Nonetheless, considering the strategic essence of design rationales, they hold the potential to assist in resolving more intricate and demanding problems.

    \item \emph{Lack of Separate Impact Analysis for Solution and Argument in APR}: While this study demonstrates the beneficial effect of design rationale on patch quality, it does not dissect and scrutinize the distinct impacts of solutions and arguments. Such an analysis is vital because more methods can identify or derive solutions from online developer discussions \cite{panthaplackel-etal-2022-learning, shi2021ispy}. Should the arguments prove to be less critical for patch generation, we could employ more techniques focused on generating solutions that could more effectively steer the patch generation process.

    \item \emph{Absence of Execution-Based Evaluation Pipeline}: While we assessed patch quality using static metrics such as CodeBLEU and full-match, these methods have notable limitations. For instance, CodeBLEU evaluates the entire function rather than focusing on the modified segments, which can result in inflated scores. Unchanged portions of the function contribute positively to the score, even though they are irrelevant to the actual repair. This skews the evaluation by overestimating the patch quality. In contrast, an execution-based approach would offer a more accurate measure of whether the patch genuinely resolves the issue by directly testing the code. Without this, our current evaluation risks missing important nuances in functional correctness and real-world applicability. Therefore, future work would benefit from integrating an execution-based evaluation pipeline to provide a more reliable and comprehensive validation of the patches.

\end{itemize}

\section{Conclusion and Future Work}
\label{conclusions}

Emulating the typical development process where solutions and their respective arguments are deliberated \emph{prior} to crafting patches, we introduce \textit{DRCodePilot}, a design rationale-driven program repair approach leveraging the capabilities of the advanced GPT-4. Besides, we devised a feedback-based self-reflective framework to prompt GPT-4 to revisit and refine its proposed fixes according to the given reference patches and suggestions for identifier replacement. For assessment, we curated a benchmark comprising 938 matched issue-patch pairs. Our experiments reveal that \textit{DRCodePilot} achieves up to 4.7x and 3.6x more full-match patches than the leading GPT-4 in our benchmark. Furthermore, we illustrate the constructive influence of design rationales, reference patches, and identifier recommendations. Moving forward, we plan to improve upon extracting design rationales and to incorporate retrieval-augmented generation techniques to tackle increasingly complex issues.

\begin{acks}
Funding for this work has been provided by the National Science Foundation of China Grant NO. 62102014 and 62177003. It is also partially supported by State Key Laboratory of Complex \& Critical Software Environment NO.SKLSDE-2023ZX-03.
\end{acks}

\bibliographystyle{ACM-Reference-Format}
\bibliography{reference}


\end{document}